\documentclass[conference]{./IEEEtran}

\renewcommand{\baselinestretch}{0.8165}
\newcommand{\subparagraph}{}

\usepackage{paralist}

\usepackage{color}
\usepackage{float}

\usepackage{enumitem}

\usepackage{graphicx}
\graphicspath{{images/}}

\usepackage[cmex10]{amsmath}

\usepackage[binary-units=true]{siunitx}

\begin{document}
\title{A Transprecision Floating-Point Platform\\for Ultra-Low Power Computing}

\author{\IEEEauthorblockN{Giuseppe~Tagliavini\IEEEauthorrefmark{1},
        Stefan~Mach\IEEEauthorrefmark{2},
        Davide~Rossi\IEEEauthorrefmark{1},
        Andrea~Marongiu\IEEEauthorrefmark{3},
        and Luca~Benini\IEEEauthorrefmark{1}\IEEEauthorrefmark{2}}
\IEEEauthorblockA{\IEEEauthorrefmark{1} DEI, University of Bologna, Italy / Email: \{giuseppe.tagliavini, davide.rossi, luca.benini\}@unibo.it}
\IEEEauthorblockA{\IEEEauthorrefmark{2} IIS, ETH Zurich, Switzerland / Email: \{smach, luca.benini\}@iis.ee.ethz.ch}
\IEEEauthorblockA{\IEEEauthorrefmark{3} DISI, University of Bologna, Italy / Email: \{a.marongiu\}@unibo.it}}


\maketitle

\begin{abstract}
In modern low-power embedded platforms, the execution of floating-point (FP) operations emerges as a major contributor to the energy consumption of compute-intensive applications with large dynamic range. Experimental evidence shows that 50\% of the energy consumed by a core and its data memory is related to FP computations.
The adoption of FP formats requiring a lower number of bits is an interesting opportunity to reduce energy consumption, since it allows to simplify the arithmetic circuitry and to reduce the memory bandwidth required to transfer data between memory and registers by enabling vectorization.
From a theoretical point of view, the adoption of multiple FP types perfectly fits with the principle of transprecision computing, allowing fine-grained control of approximation while meeting specified constraints on the precision of final results.
In this paper we propose an extended FP type system with complete hardware support to enable transprecision computing on low-power embedded processors, including two standard formats (binary32 and binary16) and two new formats (binary8 and binary16alt).
First, we introduce a software library that enables exploration of FP types by tuning both precision and dynamic range of program variables.
Then, we present a methodology to integrate our library with an external tool for precision tuning, and experimental results that highlight the clear benefits of introducing the new formats.
Finally, we present the design of a transprecision FP unit capable of handling 8-bit and 16-bit operations in addition to standard 32-bit operations.
Experimental results on FP-intensive benchmarks show that up to 90\% of FP operations can be safely scaled down to 8-bit or 16-bit formats.
Thanks to precision tuning and vectorization, execution time is decreased by 12\% and memory accesses are reduced by 27\% on average, leading to a reduction of energy consumption up to 30\%.
\end{abstract}

\section{Introduction}
\label{sec:introduction}
Nowadays most embedded applications involving numerical computations with large dynamic range are performed using \emph{binary64} (double-precision) or \emph{binary32} (single-precision) floating-point (FP) formats, described by the IEEE 754 standard \cite{zuras2008ieee}.
In these applications, the execution of FP operations emerges as a major contributor to the energy consumption.
To provide experimental evidence of this insight, we have executed a set of FP-intensive applications on PULPino \cite{gautschi2017near}, an open-source ULP microcontroller.
Results show that 30\% of the energy consumption of the core is actually due to FP operations.
Moreover, an additional 20\% is spent in moving FP operands from data memory to registers and vice versa.

To provide a compromise between energy cost and dynamic range, IEEE 754 introduces a 16-bit format referred to as \emph{binary16} (half-precision).
The introduction of \textit{binary16} represents a first step to increase the energy efficiency of FP computations, but software development flows for ULP systems still lack a methodology to evaluate the effect of reduced-precision FP variables on application requirements.
In practice, programmers often use the maximum precision provided by target platforms, following the most conservative principle: guaranteeing numerical precision of each elementary step also guarantees the precision of final results.

In recent years, significant advances in the field of approximate computing have been made, aimed at relaxing this precise-computing abstraction \cite{bekas2012low} \cite{klavik2014changing} \cite{ho2017efficient} \cite{chiang2017rigorous}.
The most promising research trends are stepping beyond of the concept of approximation itself, towards a novel paradigm -- \emph{transprecision computing} -- in which rather than \emph{tolerating} errors implied by imprecise HW or SW computations, systems are explicitly designed to deliver just the required precision for intermediate computations.
In other words, the specified constraints on the precision of final results are always met (i.e., results are not generically ``approximated'', they match a required precision), but intermediate operations can be deployed to custom, lower-precision compute units to save energy.

In this paper we propose an extended FP type system with complete hardware support to enable transprecision computing on ULP embedded platforms.
We propose the introduction of two additional formats, namely \emph{binary8} and \emph{binary16alt}.
Specifically, \emph{binary8} is a 8-bit format with low precision (3-bit mantissa), while \emph{binary16alt} is a 16-bit format complementary to the IEEE one and featuring a higher dynamic range (8-bit exponent).
To assess the benefits of this extended FP type system, we performed a precision analysis supported by additional considerations on the hardware design.
As a first step we designed a C++ library to explore the effects on application behavior when varying dynamic range and precision of program variables.
Then we adapted a set of applications representative of FP-intensive computations in the embedded domain, adopting emulated FP types and providing an interface with an external tool for precision analysis \cite{ho2017efficient}.

Our results shows that the introduction of \textit{binary8} guarantees the best trade-off between precision and dynamic range for applications that match minimum precision requirements.
Moreover, the introduction of \textit{binary16alt} extends (up to 50\%) the number of variables that can be safety scaled from a 32-bit representation to a 16-bit one.
To provide support at hardware level, we designed a dedicated \emph{transprecision FP unit}.
Our design also enables vectorial operations on smaller-than-32-bit formats, further increasing energy efficiency and performance of the core and reducing pressure on data memory.

To summarize, the main contributions of this work are (i) the introduction of a software library that enables explorations of FP types, (ii) a methodology to integrate our library with external tools for precision tuning, and (iii) an energy-efficient hardware design supporting multiple FP types.
Experimental results show that up to 90\% of FP operations can be safely scaled down to 8-bit or 16-bit formats.
Thanks to precision tuning and vectorization, execution time is decreased by 12\% and memory accesses are reduced by 27\% on average. As a major outcome, energy consumption is reduced up to 30\%.

\section{Related work}
\label{sec:related}
To overcome the limitations of fixed-format FP types, researchers have proposed multiple-precision arithmetic libraries that perform calculations on numbers with arbitrary precision.
ARPREC \cite{bailey2002arprec} and MPFR \cite{fousse2007mpfr} are two widely used libraries which provide support to multiple-precision arithmetic.
These libraries are mainly used in contexts where a high-dynamic range is required and higher computation time is considered an unavoidable side-effect, such as scientific computing.
However they are not suitable to perform explorations of less-than-32-bit FP types, since they represent exponents using a full machine word.
This approach totally prevents to simulate the behavior of FP types with a reduced number of bits, since tuning of dynamic range is not possible.
%
\textit{SoftFloat} \cite{hauser1996handling} is a library that implements standard IEEE formats, enabling a bit-accurate emulation of the FP operations performed by FP hardware units. 
\textit{Softfloat} can be easily extended to support additional formats, including the ones introduced in this work.
However program executions are extremely slow, since the library executes all the computations in software.
Moreover any modification to a FP format requires to manually modify code in several source files.
Overall, the aforementioned solutions require a full refactoring of the source code; in some cases additional software layers have been introduced to perform this task (e.g., the \textit{Boost} interval arithmetic library \cite{bronnimann2006design}).

Many research tools are available to perform automatic or semi-automatic precision tuning of program variables.
In this paper we use \textit{DistributedSearch}, a tool provided by the \textit{fpPrecisionTuning} \cite{ho2017efficient} toolchain that finds a near-optimal solution.
Its main configuration parameter is the precision of the result, expressed as a value of signal-to-quantization-noise ratio (SQNR) that program outputs must satisfy.
This tool requires a binary version of the target program, a target output (i.e., a sequence of FP numbers that are the exact result) and a configuration file.
The configuration file should include a list of numbers, which correspond to the precision bits used for program variables.
\textit{DistributedSearch} requires that the target executable is able to read the configuration file to tune the precision of its variables accordingly.
In addition, the program must provide its output results on the standard output.
On this premises, the tool runs the program multiple times, performing a heuristic search of the minimum precision that can be associated to each variable (for a fixed input set).
A second phase performs a statistical refinement to join the precision bindings derived from different input sets.
Other tools adopt more advanced techniques but their search space is restricted to standard FP types (e.g., PROMISE \cite{graillat2016auto} and \textit{Precimonious} \cite{rubio2013precimonious}), or in other cases they are limited to the analysis of functional expressions (e.g., \textit{FPTuner} \cite{chiang2017rigorous} and PRECISA \cite{moscato2017automatic}). 
As a final consideration, all these tool do not enable the analysis of the dynamic range associated to a fixed-format FP type.

On the hardware side, several recent works proposed the design of energy-efficient FP units.
Kaul et al. \cite{kaul2012varprec} implement a variable-precision multiply-and-add FP unit supporting vectorization.
Its configurations use an 8-bit exponent field, while each operand carries a 5-bit certainty field which is processed in parallel with the exponent logic, indicating the number of bits for the mantissa.
The certainty field  is used to implement automatic precision tracking, which raises the level of precision where it does not meet specified requirements.
Considering an energy consumption of 19.4pJ/FLOP, this solution seems to perform as good as our hardware design.
However the memory overhead due to precision tracking and fixed-size exponents is relevant, since the memory transfers are a major contributor to the total energy consumption.
Moreover, applications that require 32-bit variables are very inefficient due to repeated operations at lower precision that are performed until a final retry at single precision is executed.
Tong et al. \cite{tong2000redpower} explore an iterative (digit-serial) multiplier that can be used inside a FP multiplier. Their design processes 8 bits per cycle, thus operands with up to 8-bits use one cycle, operands with up to 16-bit use 2 cycles, and finally operands up to 24 bits use 3 cycles.
Power is reduced by 66\% when using the one-cycle configuration, and by 30\% when using the two-cycles one.
Again, single-precision operations become slower and memory effects are not considered.
Rzayev et al. \cite{rzayev2017deeprecon} explore various smaller-than-32-bit formats for deep learning applications.
They introduce a 8-bit FP format that is identical to \textit{binary8}, showing that vectorization enables higher performance and reduces memory energy used per operation.
However they do not propose a mixed-precision FP type system for transprecision computing.
Gautschi et al. \cite{gautschi2017shared} propose a shared FP unit adopting the logarithmic number system (LNU), which is up to $4.1\times$ more energy efficient than standard FP unit in non-linear processing kernels.
However, LNU is a domain specific approach, and not all FP operations can be implemented.
%


\section{Floating-point types and programming flow}
\label{sec:software}
\subsection{Exploration of floating-point formats}
\label{sec:exploration}
%
Applications that operate on real-valued data most commonly use IEEE 754-compliant FP formats \cite{zuras2008ieee}.
Of the standard formats, \textit{binary32} and \textit{binary64} enjoy the most wide-spread use and are also available on general-purpose consumer platforms.
While even larger formats are commonly used for scientific computations, reducing the amount of data to process and hence the width of FP formats is more suitable for power-constrained and embedded platforms.
While smaller-than-32-bit FP formats (also called \emph{minifloat}s) have been use in computer graphics applications \cite{trompouki2016towards}, their relevance is rising with the spread of energy-constrained computing platforms, such as near-sensor processing nodes and Internet-of-things endpoints.
IEEE formats are packed as the sign bit, $e$ bits for the exponent and $m$ bits for the significand (or mantissa).
By choosing a specific format, programmers enforce a trade-off between dynamic range and precision.
The dynamic range is the ratio between the largest and smallest representable values, and it is conditioned by $e$.
Conversely, the precision is the number of digits of the represented number that are preserved in FP representation, and it is uniquely defined by $m$.

As discussed in Section~\ref{sec:related}, available tools are not flexible enough to simulate arbitrary FP formats by tuning both precision and dynamic range.
To enable exploration of arbitrary FP types, we designed a dedicated C++ library, called \emph{FlexFloat}.
This library provides a generic FP type by defining a template class (\texttt{flexfloat<e,m>}) and a set of auxiliary functions for debugging and collecting statistics.
Using \textit{FlexFloat}, all FP types used in the source files can be safely replaced with instantiations of this template class without changing any other part of the program, since class methods include operator overloading.
The template parameters include the number of bits used for the exponent ($e$) and the number of bits used for the mantissa ($m$), which must be both specified as positive integer values.
For instance, \texttt{flexfloat<7, 12>} is a FP type including the sign bit, 7 bits in the exponent field and 13 bits in the mantissa field.
The \textit{FlexFloat} library also supports the encoding of denormal numbers, infinities and not-a-number (NaN) values.
Arithmetic operations are performed converting the number representation to a native type (e.g., \texttt{double}) and then \emph{sanitizing} the result, that is adjusting exponent and mantissa to obtain the exact binary representation of the original type.
This methodology guarantees shorter execution times w.r.t. emulation approaches (e.g., \textit{SoftFloat}), and it also produces the same results of a dedicated hardware unit (i.e., precise at bit level).
An automatic cast between different template instances is not allowed, so standard arithmetic operations must involve variables of the same instance.
This design choice enables a fine-grain control on the intermediate results, since the compiler notifies an error for each operator involving a type mismatch.
Consequently, programmers can choose to match the variable types to the same template instance or to insert an explicit cast.
A constructor supporting explicit conversions is provided, and it can be used to cast a \textit{FlexFloat} variable to a different template instance (e.g., from \texttt{flexfloat<e1,m1>} to \texttt{flexfloat<e2,m2>} ).
Constructors with implicit semantics are provided for standard FP types (\texttt{float}, \texttt{double} and \texttt{long double}) to simplify the usage of FP literal values.
Vice versa, an automatic cast from a FlexFloat template instance to a standard FP type is not allowed, but it can be performed by invoking an explicit cast operator.
This feature can be used to interface sections of source code that use FlexFloat and sections that are strictly bound to standard types (e.g., a call to an external library function whose source code is not available).
The main benefits of \textit{FlexFloat} are:
\begin{compactitem}
\item it produces binaries that are fast to execute, since its computations rely on native types;
\item it reduces the debugging effort, as the compiler performs early check upon template instantiation;
\item it is quite intuitive to use, since it provides the usual infix notation for arithmetic operations;
\item it can be easily integrated with external tools, having no specific requirements w.r.t. the original source code.
\end{compactitem}

\begin{figure}[!t]
\centering
\includegraphics[width=0.95\linewidth]{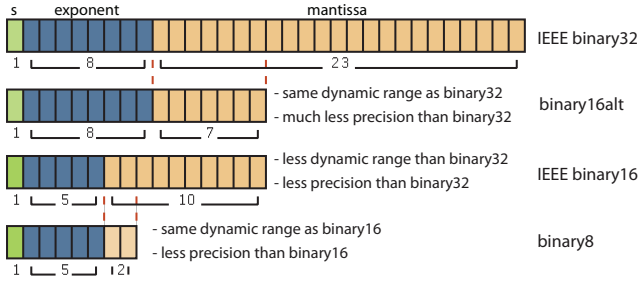}
\vspace*{-2mm}\caption{Overview of floating-point formats used throughout this work.}
\label{fig:formats}
\end{figure}
%
 resolved at compile time).
To simplify the interaction between a \textit{FlexFloat}-based program and any external tool, we designed a \emph{FlexFloat wrapper}, that is a support tool performing three steps: (i) it reads a file specifying a required precision for each program variable, then (ii) it extracts the dynamic range from a configuration file providing the map indexed by precision intervals, and finally (iii) it compiles the program sources providing the derived values for precision and dynamic range as actual parameters in the template instantiations.

To perform an exploration of FP types, we used the \textit{DistributedSearch} tool introduced in Section~\ref{sec:related}.
Since this tool performs precision tuning without considering the dynamic range of variables, we assumed a limited set of initial hypotheses to fix the dynamic range associated to specific intervals of precision bits.
Considering our target on ULP systems, we restricted our investigation to 8-bit and 16-bit formats.
Among potential 8-bit formats, we chose the mapping $(0, 3] \mapsto 5$, calling this type \emph{binary8}.
This means that any variable associated to a precision between 1 and 5 bits will be provided with an exponent of 5 bits.
This format was conceived to mirror the dynamic range of \textit{binary16} variables.
Adopting this convention, conversions between \textit{binary8} and \textit{binary16} only affect precision but do not saturate for values of large magnitudes. Additionally, operations on \textit{binary8} become very cheap in
hardware since there are only two explicit mantissa bits.
As regards 16-bit formats, we considered the mapping corresponding to \textit{binary16}, that is $(0, 11] \mapsto 5$, and an alternative mapping that we called \emph{binary16alt}, corresponding to $(0, 8] \mapsto 8$.
Basically, 8 is the number of bits used for the exponent field in \textit{binary32}, so this value is a logical upper-bound for any 16-bit format.
Again, using the same number of exponent bits of the \textit{binary32} format makes conversions much cheaper.
Figure~\ref{fig:formats} summarizes the FP formats used throughout this work.

Table~\ref{tab:experiments01} shows the results of our preliminary analysis, reporting the total number of variables associated to each type.
These values are obtained executing \textit{DistributedSearch} on our set of benchmarks constrained with a precision of $10^{-1}$.
We considered two different configurations of the FP type system, namely V1 (including \textit{binary8}, \textit{binary16}, \textit{binary32}) and V2 (adding \textit{binary16alt} to V1).
\begin{table}[!b]
\caption{Variables classified by type type using V1 and V2 type systems.}
\label{tab:experiments01}
\centering
\begin{tabular}{ | l | l | l | l | l | }
\hline
                 & \textbf{binary8} & \textbf{binary16} & \textbf{binary16alt} & \textbf{binary32} \\
\hline
\textbf{V1}      &              10  &               29  &                  -   &                72 \\
\hline
\textbf{V2}      &              19  &               10  &                  41  &                41 \\
\hline
\end{tabular}
\end{table}
%
As a first consideration, \textit{binary8} is used for 17\% of the variables in the best case.
This format is extremely beneficial in reducing the energy consumption, since it allows to simplify circuitry complexity and it enables vectorization.
It is noteworthy that supporting both 16-bit formats contribute in decreasing the number of 32-bit variables in the program w.r.t. the usage of a single 16-bit format.
A drawback of \textit{binary16} is that both dynamic range and precision are diminished when compared to \textit{binary32}. This leads to saturation when converting values with large dynamic from \textit{binary32}, disqualifying the 16-bit format from being used for transprecision tuning in these cases.
Conversely, \textit{binary16alt} features the same dynamic range as \textit{binary32}, allowing the whole range of values to be converted - albeit with a much larger granularity.
In some cases applications do not exploit the dynamic range provided by \textit{binary16alt}, and at the same time they require a higher precision, so our intuition is that we need both types.
A further evaluation is provided in Section~\ref{sec:static}.

\subsection{Transprecision programming flow}
\label{sec:flow}
\begin{figure}[!t]
\centering
\includegraphics[width=0.98\linewidth]{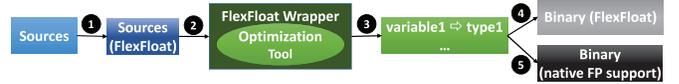}
\vspace*{-2mm}\caption{Overview of the programming flow.}
\label{fig:flow}
\end{figure}

Figure~\ref{fig:flow} depicts the transprecision programming flow that we adopted throughout this work.
As a first step, application sources are modified to replace standard FP types with multiple instances of \texttt{flexfloat<ex,mx>}, where $ex$ and $mx$ are variable-specific parameters.
Then a tool for precision tuning is invoked (step 2), and different values for $ex$ and $mx$ are explored using the \textit{FlexFloat wrapper}.
After this tuning, program variables are uniquely mapped to supported FP types (step 3).
Using this mapping, \textit{FlexFloat} can provide statistics on the number of operations and casts performed for each FP type which is instantiated (step 4).
%
Moreover, a version of FlexFloat providing explicit template specialization is provided to replace simulated operations with native ones (step 5).
This step requires that the compiler for the target platform supports all the FP types provided by the mapping.

\begin{figure*}[!ht]
\centering
\includegraphics[width=0.9\textwidth]{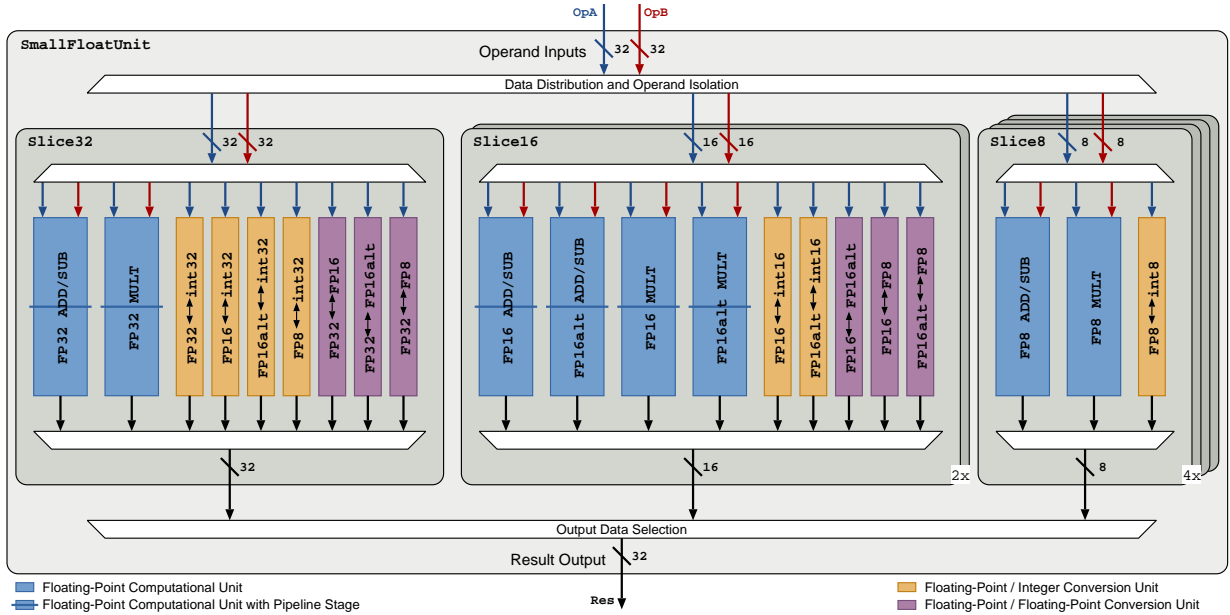}
\caption{Simplified block diagram of the designed hardware unit datapath. Control logic as well as data to and from duplicated slices is omitted.}
\label{fig:hwunit}
\end{figure*}
\section{Transprecision floating point unit}
\label{sec:hardware}
To evaluate the potential of the FP formats introduced in Section~\ref{sec:software}, we designed a transprecision FP unit supporting vectorization of reduced-precision operations.
%
The hardware unit is built up from three types of slices, each with a fixed width of 32-bit, 16-bit and 8-bit, respectively. Each slice hosts operations on the FP formats that match the slice width, as well as conversion operations.
The supported arithmetic operations are addition, subtraction and multiplication.
The conversion operations include casts to and from integers (both signed and unsigned) as well as casts among the FP formats.
Moreover, the narrower slices are replicated in order to enable sub-word parallelism inside the unit. Thus, two 16-bit or four 8-bit FP operations can be executed simultaneously.
Following the single-instruction-multiple-data (SIMD) paradigm, the proposed unit can run scalar operations when only one slice for a given precision is active, and vectorial operation when all the slices of a given precision are active.

The various individual operation blocks are instances of Synopsys DesignWare FP Datapath components.
%
As a power saving feature, the unit employs operand silencing to all unused operations and formats by forcing zero to prevent transistor switching.
To meet the timing requirements of the container core, arithmetic operations in \textit{binary32} as well as both 16-bit formats are pipelined with one stage,  hence featuring a bandwidth of one operation per cycle and a latency of two clock cycles.
Arithmetic operations in \textit{binary8} as well as all conversion operations have a one cycle latency.
Area optimization of the transprecision FPU and its integration into the core will be completed as future work.

\section{Experimental results}
\label{sec:experiments}

\subsection{Evaluation Methodology}
\label{sec:evaluation}
Experiments have been performed on a set of applications which implement key algorithms for two domains of ULP systems, near-sensor computing and embedded machine learning:
\begin{compactitem}
\item JACOBI applies the Jacobi method to a 2D heat grid:
\item KNN computes the k-nearest neighbors of an input value using euclidean distance;
\item PCA performs the principal component analysis;
\item DWT computes the discrete wavelet transform;
\item SVM is the prediction stage of a support vector machine;
\item CONV implements a $5\times5$ convolution kernel.
\end{compactitem}

Precision tuning has been performed using the \textit{fpPrecisionTuning} toolsuite on an x86 workstation, adopting the programming flow described in Section~\ref{sec:flow}.
Since sub-word vectorization is not supported by the current \textit{FlexFloat} implementation, program sections that are vectorizable are manually tagged in the source code, and the library provides a distinct report for vectorial operations and casts.
%
%
The application sources have been compiled using the GCC compiler with a RISC-V backend optimized for PULPino, which provides support for the single-precision FP type defined in the RISC-V instruction set architecture (ISA).
Binaries have been executed on the PULPino virtual platform, which is cycle accurate and provides detailed statistics.
The virtual platform reports the number of cycles required to execute each instruction that is used in the binary file, targeting the whole program or delimited code regions.
The current version of GCC does not include a set of instructions to handle \textit{binary16}, \textit{binary16alt} and \textit{binary8} formats.
Since the latency of \textit{binary16} operations is the same of \textit{binary32} ones, we have used the \textit{binary32} type to measure the exact number of cycles required by each instruction to execute.
This value depends by the ability of the compiler to schedule other classes of operations (non-FP, \textit{binary8} or casts) to fill latency cycles and avoid stalls in the core pipeline, so it is strictly dependent on both application and compiler back-end.
\textit{binary8} operations and FP casts always require a single cycle, so their contribution to execution time has been accumulated analytically.

For evaluation of the hardware architecture, the design unit was synthesized for the UMC~\SI{65}{\nano\meter} technology using worst case libraries (\SI{1.08}{\volt}, \SI{125}{\celsius}).
To have an accurate estimation of the power consumption of the transprecision FPU, we performed post-place-\&-route power simulations.
The target frequency for the post-layout design was set to \SI{350}{\mega\hertz}, using worst-case conditions.
Results take into account the switching activity of input and output registers, added at the boundaries of the unit to evaluate their performance, negligible with respect to the power of the arithmetic units themselves.
Energy costs of FP operations were obtained through simulation of the post-layout design in all modes of operation, again using worst-case conditions.
Values provided to the unit were generated in a random fashion, making sure that no invalid values were generated.
Namely, no NaN or infinity values were applied and operands were chosen sufficiently close to each other such that operand cancellation would not occur during addition or subtraction.
For conversions, only values that can be mapped to the target type were applied to eliminate over- or underflow.
This was done to simulate normal operation wherein operation is done on meaningful data, while operand cancellation or invalid inputs lead to significantly diminished switching activity inside the operational units.
The energy cost of each non-FP instruction executed by the core includes contribution of core logic, instruction memory and data memory.
Even if the transprecision FPU has not been integrated in the PULPino core yet, to collect energy measures we have considered the contribution of moving data to/from input and output registers of the FP unit, and also the cost of idle cycles due to pipeline stalls (for both 16-bit and 32-bit instructions).

\subsection{Precision Tuning}
\label{sec:static}
\begin{figure}
\centering
\includegraphics[width=0.95\linewidth]{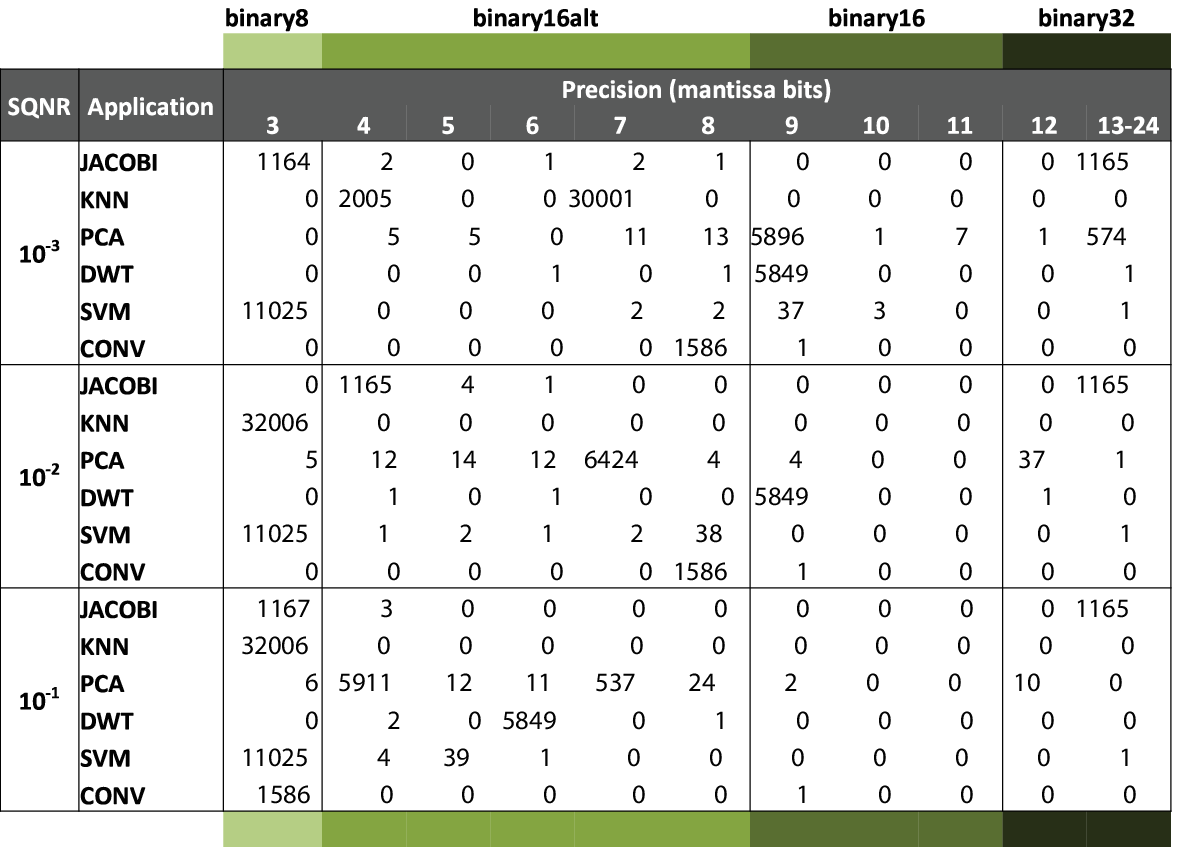}
\vspace*{-2mm}\caption{Precision tuning of program variables for three precision requirements.}
\label{fig:exp01}
\end{figure}
%
The table in Figure~\ref{fig:exp01} shows the results of the precision tuning, performed for three precision requirements (SQNR = $10^{-3}$, $10^{-2}$, $10^{-1}$).
Rows correspond to applications and columns to precision bits.
The reported values represent the number of memory locations (scalar variables or array elements) requiring the minimum number of bits of their column to meet precision constraints.
Color bands show the mapping between precision bits and the FP type system introduced in Section~\ref{sec:software}.
KNN and SVM make wide use of \textit{binary8} data, while in general other application do not.
In fact \textit{binary8} emerges a format which is profitable in specific applications domains, while \textit{binary16} is a good candidate for a wider use.
Moreover, it is evident that most elements in the interval $[9,11]$ are concentrated in column $9$, which is the minimum number of precision bits required for a \textit{binary16} type.
This is due to the fact that these elements strictly require the additional precision provided by \textit{binary16} w.r.t. \textit{binary16alt}, which means that both types are useful in different contexts.
For the same reason, variables in column $4$ are more than variables in column $5$, since they include all cases that do not require a wider dynamic range w.r.t. \textit{binary8} (regardless of the precision).
Conversely, variables that require high precision usually require more than 12 precision bits, and they are concentrated in the last column.

\subsection{Execution time and memory accesses}
\label{sec:dynamic}
\begin{figure}
\centering
\includegraphics[width=0.95\linewidth]{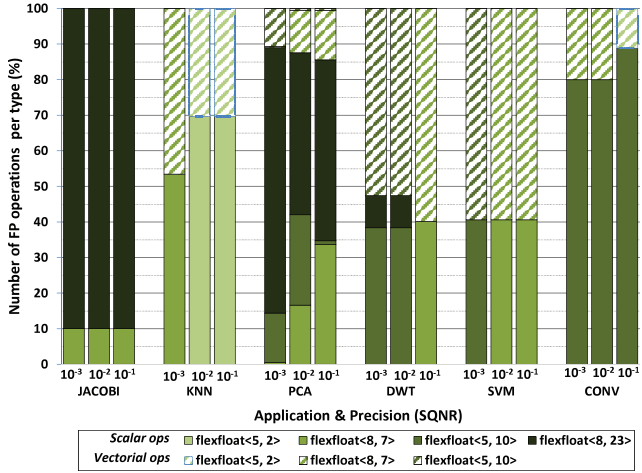}
\vspace*{-2mm}\caption{Breakdown of FP operations for three precision requirements.}
\label{fig:exp02}
\end{figure}
\begin{figure*}[!ht]
\centering
\includegraphics[width=0.95\linewidth]{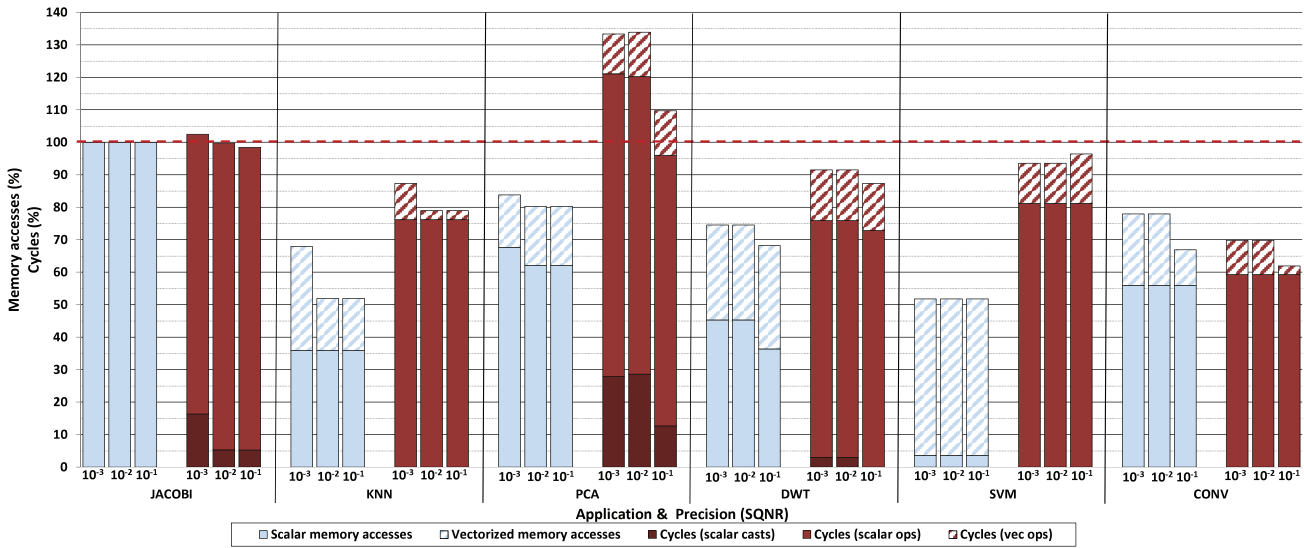}
\vspace*{-2mm}\caption{Memory access and cycles for three precision requirements, normalized to \textit{binary32} baseline.}
\label{fig:exp03}
\end{figure*}
%
Figure~\ref{fig:exp02} shows a breakdown of the FP operations performed by each application, taking into account the same precision requirements of previous section.
This is a dynamic view of the FP type system at run-time (whereas Figure~\ref{fig:exp01} provides a static view after precision tuning).
Each bar segment quantifies the contribution of a specific type to the total number of FP operations, discriminating scalar and vectorial operations.
In JACOBI and PCA there is a major contribution of 32-bit operations, which is a first trait that adversely affects a potential reduction of energy consumption.
Another negative aspect is the lack of vectorial operations, that is pathological in JACOBI.
In this work we have not considered any advanced coding techniques (e.g., manual code vectorization), but we have based out analysis on off-the-shelf versions of applications that could be further optimized to achieve following the guidelines derived from our considerations.

Figure~\ref{fig:exp03} depicts two groups of bars for each application, reporting memory accesses and executions cycles.
Values are normalized to the \textit{binary32} version of the application, that acts as a baseline.
Vectorial memory accesses, cycles spent in vectorial operations and cycle spent in cast operations are highlighted with a different pattern.
As showed in the previous section, JACOBI does not perform any vectorial operation.
Moreover the number of cycles is equivalent to the original version, since this application only uses a limited number of \textit{binary16alt} variables without exploiting vectorization.
In the most general case, the number of cycles can even exceed the baseline, since cast operations between different FP types are introduced (e.g., JACOBI when SQNR = $10^{-3}$).
As a major limitation, current tools for precision tuning do not take into account the cost of casts, since they aim at minimizing the number of precision bits used by any variable and no additional optimization goal can be specified.
This effect is further exacerbated in PCA, where the number of casts required after the tuning process exceeds 10\% (SQNR = $10^{-1}$) and 20\% (SQNR = $10^{-2}$ and $10^{-3}$)).
In other benchmarks we can observe evident benefits in memory accesses and cycles, mainly due to vectorization, while the overhead of cast operations is not relevant.
%
SVM shows the maximum reduction of memory accesses, that is 48\%, since 60\% of FP operations are vectorizable (for any precision requirement).
On average, the execution time is decreased by 12\% and memory accesses are reduced by 27\%.
Considering JACOBI and PCA as outliers, these values turn into 17\% and 36\%.

\subsection{Energy consumption}
\label{sec:energy}
\begin{figure}
\centering
\includegraphics[width=0.95\linewidth]{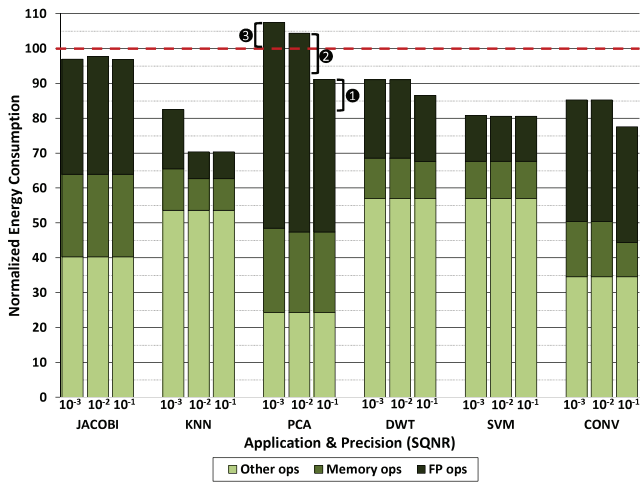}
\vspace*{-2mm}\caption{Energy consumption normalized to \textit{binary32} baseline.}
\label{fig:exp04}
\end{figure}
%
Figure~\ref{fig:exp04} shows the energy consumption of each application, normalized to the \textit{binary32} baseline.
Each bar contains three contributions, the FP operations (FP ops), the memory accesses (Memory ops) and all the other instructions that are executed by the core (Other ops).
These numbers can be easily justified by the considerations in the previous section.
On average, the energy consumption of JACOBI is 97\%, since this application makes limited use of smaller-than-32-bit types and does not exploit the benefits of vectorization.
The energy consumption of PCA is 7\% and 8\% greater than the baseline for two precision requirements (SQNR = $10^{-3}$ and $10^{-2}$), due to the high number of casts coupled with a predominant number of scalar operations on \textit{binary32} values.
The other applications have average energy savings around  18\% compared to the baseline, with a maximum of 30\% measured for KNN.
Considering the results of Section~\ref{sec:static}, the behavior of KNN is related to three main characteristics, (i) it uses the \textit{binary8} type for all program variables, (ii) it exploits vectorization, and finally  (iii) it requires a limited number of non-vectorized memory accesses.

Advanced vectorization techniques can provide huge benefits whenever an application provides a relevant percentage of smaller-than-32-bit operations after the tuning process.
To demonstrate this assumption, we applied manual vectorization to PCA, thus reducing the energy consumption to lower values (101\%, 96\% and 85\%).
These gains are marked on Figure~\ref{fig:exp04} by labels 1, 2 and 3.
Further energy savings can be only achieved by reducing the contribution of casts with the support of smarter tools for precision tuning.

\section{Conclusion}
\label{sec:conclusion}
%
This work introduces an extended FP type system with complete hardware support to enable transprecision computing on ULP embedded platforms.
Experimental results show that our approach is effective in reducing energy consumption by leveraging the knobs provided by the extended FP type system and thanks to vectorization support.
The energy consumption is reduced on average by 18\%, and up to 30\% for specific applications.
At the same time, execution time is decreased by 12\% and memory accesses are reduced by 27\%.

Our future work will be focused on three main aspects.
First, the study of new techniques of precision tuning, that take into account the costs of casts with the aim to formulate a multi-objective optimization problem.
Second, the optimization of transprecision hardware units, to achieve better performance and minimize the area.
Third, the investigation of compiler passes and vectorization techniques to better exploit transprecision opportunities at compile time.

\section*{Acknowledgment}
This work has been partially supported by the European FP7 ERC project MULTITHERMAN (g.a. 291125) and by the European H2020 FET project OPRECOMP (g.a. 732631).


\bibliographystyle{DATE18-FPext}
\bstctlcite{IEEEexample:BSTcontrol}
\footnotesize{\bibliography{IEEEabrv,references}}

\end{document}